\documentclass[aps,pra,notitlepage,onecolumn,groupedaddress,longbibliography,amssymb,floatfix]{revtex4-1}

\usepackage{amsmath}
\usepackage{amsfonts}
\usepackage{graphicx}
\usepackage{mathptmx}
\usepackage{bm}
\usepackage{enumerate}
\usepackage{subfigure}
\usepackage{color}
\usepackage{listings}
\usepackage{hyperref}

\newcommand{\ket}[1]{| #1 \rangle}
\newcommand{\bra}[1]{\langle #1 |}

\begin{document}

\title{Discrete-event simulation of quantum walks}

\author{M. Willsch}
\affiliation{J\"ulich Supercomputing Centre, Institute for Advanced Simulation,\\
Forschungszentrum J\"ulich, D-52425 J\"ulich, Germany}
\author{D. Willsch}
\affiliation{J\"ulich Supercomputing Centre, Institute for Advanced Simulation,\\
Forschungszentrum J\"ulich, D-52425 J\"ulich, Germany}
\author{K. Michielsen}
\affiliation{J\"ulich Supercomputing Centre, Institute for Advanced Simulation,\\
Forschungszentrum J\"ulich, D-52425 J\"ulich, Germany}
\affiliation{RWTH Aachen University, D-52056 Aachen, Germany}
\author{H. De Raedt}
\affiliation{J\"ulich Supercomputing Centre, Institute for Advanced Simulation,\\
Forschungszentrum J\"ulich, D-52425 J\"ulich, Germany}
\affiliation{Zernike Institute for Advanced Materials,\\
University of Groningen, Nijenborgh 4, NL-9747 AG Groningen, The Netherlands}

\date{\today}

\begin{abstract}
We use discrete-event simulation on a digital computer to study
two different models of experimentally realizable quantum walks.
The simulation models comply with Einstein locality,
are as ``realistic'' as the one of the simple random walk in that
the particles follow well-defined trajectories,
are void of concepts such as particle-wave duality and wave-function collapse,
and reproduce the quantum-theoretical results by means of a cause-and-effect, event-by-event process.
Our simulation model for the quantum walk experiment presented in
[C. Robens et al., Phys. Rev. X 5, 011003 (2015)]
reproduces the result of that experiment.
Therefore, the claim that the result of the experiment ``rigorously excludes (i.e., falsifies)
any explanation of quantum transport based on classical, well-defined trajectories''
needs to be revised.
\end{abstract}

\maketitle

\section{Introduction}\label{INT}

A particle is said to perform a simple random walk (SRW) over a set of lattice points
(enumerated by integers) when at each time step, it jumps to one of its neighboring points, and
the direction of the jump is determined by a random variable~\cite{PEAR05,GRIM01}.
Random walks find applications in many diverse fields, too many to list them here.

The term ``quantum random walk'' was introduced in 1993~\cite{AHAR93}
and emphasizes the analogy to the simple random walk on a lattice.
However, the time evolution of a ``quantum random walk'' is deterministic and reversible~\cite{KNIG03},
not random at all, so the term \emph{quantum walk} (QW) is more apt.
There are various kinds of proposals and implementations of QWs using optical lattices~\cite{DURR02,KARS09,ROBE15},
ion traps~\cite{TRAV02,SCHM09,ZAHR10}, microwave cavities
\cite{SAND03}, or optical networks~\cite{JEON04,SCHR10,BROO10}.
A review covering various aspects of QWs is given in Ref.~\onlinecite{KEMP03}.

The basic idea of the QW is similar to that of the SRW.
Instead of using a random variable to decide which way to jump,
an internal degree of freedom (e.g. spin or polarization)
is used to determine the direction of the jump.
This internal degree of freedom changes its state according to the rules of quantum theory,
that is by a unitary transformation.

For simplicity, in this paper, we consider the case where
this state is described by a 2-dimensional Hilbert space
(e.g. spin up $|{\uparrow}\rangle$ and spin down $|{\downarrow}\rangle$)
and the particle makes nearest-neighbor hops on a one-dimensional lattice.
Compared to the SRW, the new feature is that at each jump,
the state of the spin changes by a unitary transformation, e.g. a Hadamard transformation.
The particle moves to the right if the projection of the spin (along the $z$-axis by convention)
is up $|{\uparrow}\rangle$ and moves to the left if its spin is down $|{\downarrow}\rangle$.

In symbols, this process is formalized as follows.
The basis states of the Hilbert space are $|x,s\rangle$, where
$x\in\{-L,\ldots,L\}$ labels the position on the one-dimensional lattice of $X=2L+1$ sites, and
$s\in\{{\uparrow},{\downarrow}\}$ labels the eigenstates of the $z$-component
of the Pauli matrices describing the internal degree of freedom.
In terms of the  basis states, the wave function at step $l$ reads
\begin{equation}
|\Phi^{(l)}\rangle=\sum_{x=-L}^{L}
\phi_{x,{\uparrow}}^{(l)} |x,{\uparrow}\rangle
+\phi_{x,{\downarrow}}^{(l)} |x,{\downarrow}\rangle
\quad,\quad \sum_{x=-L}^{L} \left|\phi_{x,{\uparrow}}^{(l)}\right|^2+\left|\phi_{x,{\downarrow}}^{(l)}\right|^2=1
,
\label{INT0}
\end{equation}
and is related to the initial state $|\Phi^{(0)}\rangle=|0,{\uparrow}\rangle$ by
\begin{equation}
|\Phi^{(l)}\rangle=\left(S H\right)^l|\Phi^{(0)}\rangle
,
\label{INT1}
\end{equation}
where
\begin{equation}
S=\sum_{x=-L+1}^{L} \ket{x-1,{\uparrow}}\bra{x,{\uparrow}} + \sum_{x=-L}^{L-1} \ket{x+1,{\downarrow}}\bra{x,{\downarrow}}
\label{INT2}
\end{equation}
is the operator that implements the particle jump,
\begin{equation}
\ket{{\uparrow}}\bra{{\uparrow}}=
\begin{pmatrix}1 & 0 \\0 &0\end{pmatrix}
\;,\quad
\ket{{\downarrow}}\bra{{\downarrow}}=
\begin{pmatrix}0 & 0 \\0 &1\end{pmatrix}
\;,\quad
\ket{{\uparrow}}\bra{{\downarrow}}=
\begin{pmatrix}0 & 1 \\0 &0\end{pmatrix}
\;,\quad
\ket{{\downarrow}}\bra{{\uparrow}}=
\begin{pmatrix}0 & 1 \\0 &0\end{pmatrix}
\label{INT2a}
\end{equation}
are the spin projection operators,
and
\begin{equation}
   H=\frac{1}{\sqrt{2}}\begin{pmatrix}
                          1 & \phantom{-}1 \\
                          1 &-1
                       \end{pmatrix}
\label{INT3}
\end{equation}
is the Hadamard operation, acting on the spin degree-of-freedom only.
We only consider the case that the number of steps is smaller than or equal to $L$,
meaning that the particle initially localized at $x=0$ never goes beyond the boundaries of the lattice.

QWs are different from SRWs in that
the latter cannot display interference phenomena whereas the former, being described
in terms of the evolution of a wave function, can.
In addition, the probability distribution of a QW (starting from the initial state
$|\Phi^{(0)}\rangle=|0,{\uparrow}\rangle$) is not necessarily symmetric w.r.t. $x=0$,
unlike the probability distribution of a SRW for a particle initially at $x=0$.
Furthermore, the variance of $x$ is nonlinear in the number of steps $L$~\cite{KEMP03}.

There are two distinct views of the formulation of the QW.
The first uses the particle picture to spell out the rules by which a particle changes its position and spin.
Although the spin is often regarded as a characteristic quantum feature, if there is only one spin in play,
we can equally well represent this spin by a unit vector on a Bloch sphere, a genuine classical-mechanical construct.
The quantization of the spin only enters through the digitalization of its projection on the $z$-axis,
a process very similar to the tossing of a coin, which during its flight usually rotates.
This pictorial description of the motion of a {\bf single} particle is as ``realistic'' as the one of the SRW.
Indeed, at any time the particle is at a definite position and the measurement
of the internal degree of freedom yields
an unpredictable outcome (the mapping of the unit vector to ``spin-up'' or ``spin-down''),
determining the direction of the jump.

In the second view of the formulation of the QW, use of wave mechanics is made
in order to describe the evolution of a collection of particles,
prepared in the same initial state (position and spin).
The realistic view is lost when we impose that the time evolution of a single particle and its
internal degree of freedom are to be described in terms of a wave function that evolves in time according to the rules
of quantum theory, Eq.~(\ref{INT1}) in the case at hand.

\section{Aim of the paper}
In this paper, we demonstrate that QWs can be modeled without ever having to resort to the notion
of particle-wave duality, the wave function of the particle, the update rule Eq.~(\ref{INT1}), etc.
Specifically, we show that it is possible to construct a discrete-event simulation (DES) that
is as realistic as the model of the SRW, complies with Einstein's notion of local causality~\cite{HESS19},
and reproduces the results of quantum theory without using expressions such as Eqs.~(\ref{INT0}) or~(\ref{INT1}).
In this respect, DES constitutes a ``subquantum'' model that agrees with the statistical results of quantum theory but additionally gives a description in terms of individual events in contrast to quantum theory which only gives collective, statistical predictions.

DES is a general methodology for simulating the time evolution of a system as a discrete sequence of consecutive events.
In the application at hand, there are four different kinds of events, namely
a particle starting its walk,
an operation acting on the second degree of freedom (e.g.\ the spin) of a particle,
a particle moving from one lattice site to the next according to the state of the second degree of freedom,
and a particle being counted and removed by detectors positioned at each of the lattice sites and activated after a particle made the maximum number of allowed jumps.

Simulation of a SRW is one of the simplest applications of DES.
In the DES of both the SRW and the QW, each walker follows a well-defined trajectory but in contrast
to the former, the latter yields distributions of particles over the lattice which agree with
the quantum-theoretical prediction, not with a distribution originating from a diffusion process.

We also use DES to reproduce the experimental data of a particular QW experiment with atoms~\cite{ROBE15},
which ``rigorously excludes (i.e., falsifies)
any explanation of quantum transport based on classical, well-defined trajectories'',
in blatant contradiction with the fact that each of the particles in the DES follows a well-defined trajectory and
the DES reproduces the experimental data.
In particular, we show that the DES can produce data that either violates or does not violate the
Leggett-Garg inequality (LGI)~\cite{LEGG85}, depending on the treatment of the data~\cite{NIEU11,HESS15}.
This implies that the QW by itself is not the cause of a violation of the LGI.
Note that in DES it is trivial to perform non-invasive measurements, an essential requirement
for the application of the LGI~\cite{LEGG85}.

\section{Discrete-event simulation}
For our demonstration, we build on the DES approach introduced in Ref.~\onlinecite{RAED05b},
which reproduces the experimental and quantum-theoretical results of
many fundamental quantum-physics experiments with photons and neutrons~\cite{MICH14a}.
In essence, we use DES on a digital computer as a metaphor for a perfect laboratory experiment~\cite{RAED16c}.
A salient feature of any DES implementation is that all variables which enter the model have definite values
and are known at all times.
The application of DES to the QW is based on the following ideas:

\begin{enumerate}[(a)]
\item
The moving object is treated as a particle carrying a unit vector and making nearest-neighbor jumps on a one-dimensional lattice.
\item
There are ``processing units'' which can be thought of as being placed on the lattice sites.
Depending on the unit vector that the particle carries when it enters a processing unit,
the latter may rotate the unit vector and tell the particle where to jump to.
\item
Each particle can only take one definite path.
In this sense, our DES of a QW is as ``realistic'' as the DES of a SRW.
\item
A particle can arrive at only one detector.
The function of the detector is to count the particle and to remove it from the lattice.
Each detection event is caused by exactly one particle making a walk.
Of course, being a simulation on a digital computer, during the DES, the position of
the particle and its unit vector can be ``read out'' at any time, without disturbing anything.
\item
A particle is not allowed to start its walk as long as there is another particle present on the lattice,
implying that there can be no direct interaction between particles.
\item
Interference results from the adaptive dynamics of the processing units. In the case at hand, a processing unit models a beam splitter with two input and two output ports (see below).
Input to such a processing unit are the port at which the particle enters and the orientation of the unit vector.
The adaptive dynamics changes the internal state of the processing unit in a deterministic manner.
The internal state determines the output port by which a particle leaves the processing unit.\label{item6}
\item
After processing many particles (100000 in the case at hand), the relative frequencies of the detector counts
agree with the probabilities obtained from the quantum-theoretical description.
\end{enumerate}

In a DES, we can read off, at any time, the value of a physical quantity without
changing the state of the system and we explicitly exclude from consideration DES implementations
that violate Einstein's criterion of local causality.
Specifically, our DES models satisfy the locality criteria of category 0,
as defined in Ref.~\onlinecite{HESS19}, that is they are void of interactions
(such as those appearing in the hydrodynamic/Bohm interpretation of quantum theory~\cite{MADE27,BOHM52})
that violate Einstein's criterion of local causality.
In summary, our DES approach satisfies the criteria for a local realist model.

Our DES is manifestly ``non-quantum mechanical'' in the sense that
there is no wave function describing the state of the particle in space-time but instead there are definite
particle trajectories. Still the rules by which these trajectories are formed
cannot be described by ``Newtonian mechanics''.
Clearly, without calling upon magic,
one cannot have individual particles following well-defined trajectories interfere unless there is a
mechanism at work that provides some form of indirect communication between successive particles starting their walk.
As mentioned in (\ref{item6}) above, in our DES approach, this indirect communication is the result of the adaptive (non-Newtonian)
dynamics of the processing unit.

At this point of the discussion, we wish to draw attention to a paper of W. Duane~\cite{DUAN23}.
Duane proposed that, in addition to the quantum rules for energy and angular momentum,
there is a similar rule for the linear momentum and then showed that with this rule one can explain
the diffraction of X-rays from a crystal without reference to interference of waves~\cite{DUAN23}.
In plain words, the key point of Duane's work may be formulated as follows:
there is no reason to attach a wave function to a particle if there is plenty of wave-like motion in the crystal
with which the X-rays interact.
At the time of the development of quantum theory,
the latter experiment was generally taken as strong evidence for the dual particle-wave character.
An extensive discussion of the negative impact of the particle-wave duality and the development of a deeper understanding
of where quantum theory comes from and what it entails is given by A. Land\'e in a
series of papers~\cite{LAND65,LAND66,LAND69,LAND75} and a book~\cite{LAND15}.

We mention Duane's work~\cite{DUAN23} here
because in essence, a similar idea is also used
to construct the rules of operation for the processing units in our DES.
Indeed, a quick glance at the structure of the DES algorithm for a beam splitter~\cite{RAED05b,MICH14a} shows that
the internal state of this unit is represented by a real-valued vector
of length two and a complex-valued vector of length four.
The decision about the port at which the particle leaves the beam splitter
involves the combination of these two vectors and a multiplication
by a $4\times4$ unitary matrix.
In other words, we have
attached a kind of ``wave function'' to the material (of the beam splitter),
meaning that this ``wave function'' is local to the device.
In the case of a beam splitter for light, the internal state, the ``wave function'',
is just another word for the electrical polarization vector of the material~\cite{MICH14a}
and has little relation to the particle wave function that appears in quantum theory.
An essential ingredient of the processing unit, its capability
of adapting (learning) its state from the particles that it receives
on its input ports, as well as the rule to send particles out,
cannot be inferred from the work of Duane.
They are designed such that the DES is able to reproduce,
event-by-event (particle-by-particle), in a cause-and-effect manner, the
values of the probabilities predicted by quantum theory~\cite{RAED05b,MICH14a}.

\begin{figure}[th]
\centering
\includegraphics[width=0.65\hsize]{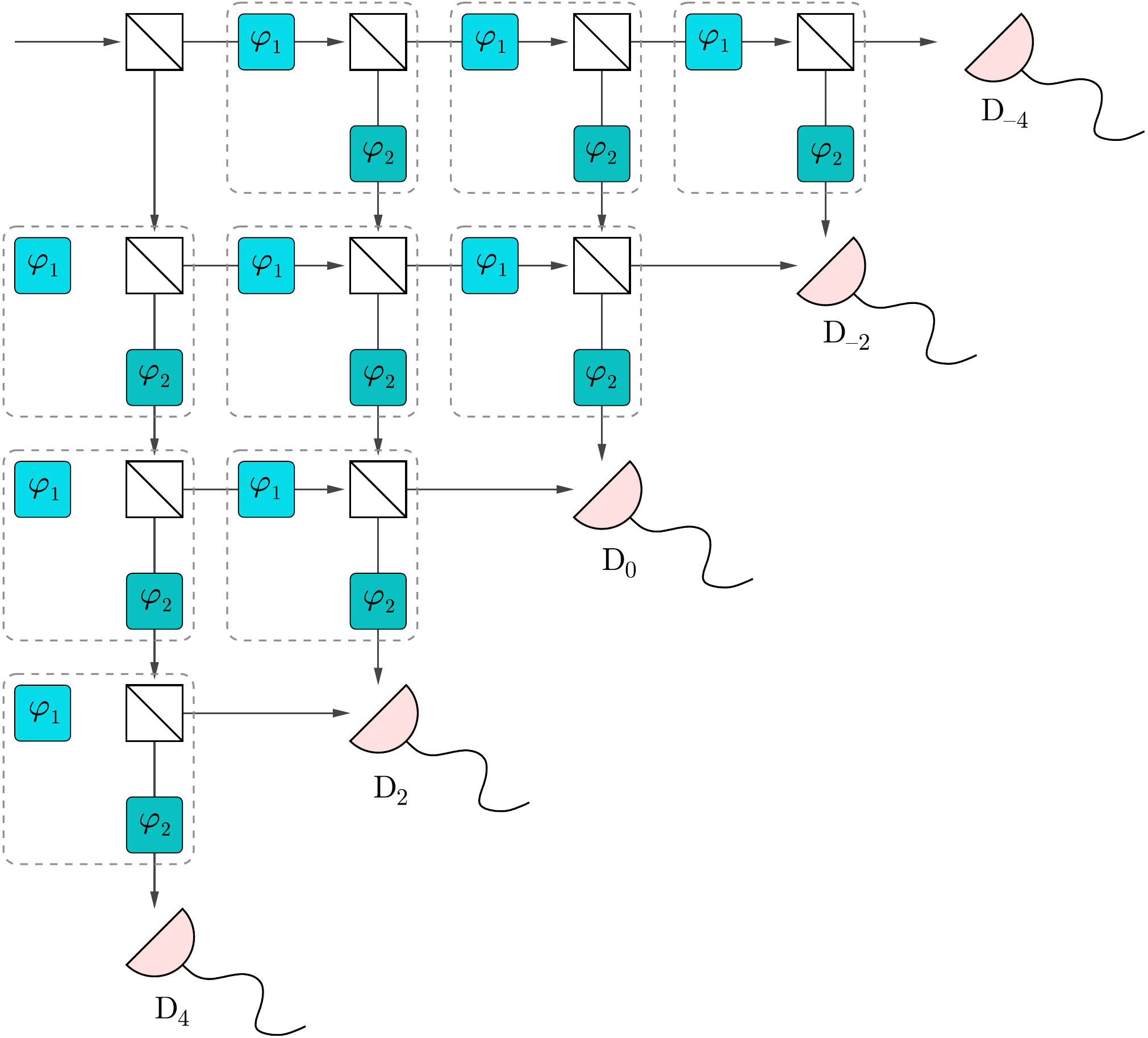}
\caption{Setup of a realization of a QW experiment~\cite{JEON04}
on a line of $L=4$ levels ($X=9$ lattice sites).
The solid (cyan) boxes represent phase shifters, shifting the phase of the wave
by angles $\varphi_1$ or $\varphi_2$, respectively.
Open squares with a diagonal line represent 50:50 beam splitters.
Half circles (pink) with a tail denote detectors at level 4, placed at lattice sites $x=-4,-2,0,2,4$.
Each group of three processing units, marked by a dashed border, causes the particle to jump left or right.}
   \label{fig1}
\end{figure}

\section{Discrete-event simulation of a QW}\label{QW}
In this section, we present the results of a DES for a QW on a line which can
be implemented by a network of beam splitters, phase shifters and photodetectors~\cite{JEON04}.
An interesting point of this implementation is that light waves can be used to simulate the QW,
i.e. we can use Maxwell's equations for electromagnetic waves to simulate a quantum system.
Of course, this is not really a surprise as the description of beam splitters, phase shifters etc.,
uses Jones-vector calculus which is, in essence, the same as the quantum-theoretical
description in terms of Eqs.~(\ref{INT0})--(\ref{INT3}).
As explained earlier, the main point of performing a DES for a QW is that it uses an event-by-event, particle-based
approach that is as realistic as the description of a SRW and does not rely on the quantum
formalism embodied in Eqs.~(\ref{INT0})--(\ref{INT3}).

The layout of the proposed experiment is shown in Fig.~\ref{fig1}.
The function of the beam splitters is to create the superposition of the two input modes.
In Jones-vector calculus or quantum theory (see Appendix~\ref{QT}), the matrix describing the operation
of a beam splitter is given by
\begin{align}
\label{MBS}
M_\mathrm{BS}=\frac{1}{\sqrt{2}}\begin{pmatrix} 1 & i \\ i & 1
\end{pmatrix}.
\end{align}
Two phase shifters, with their Jones matrix representation given by
\begin{align}
\label{MPhi1Phi2}
  M_{\varphi_1}= \begin{pmatrix}
      e^{i\varphi_1} & 0 \\ 0 & 1
   \end{pmatrix}
\quad\text{and}\quad M_{\varphi_2}=\begin{pmatrix}
              1 & 0 \\ 0 & e^{i\varphi_2}
           \end{pmatrix},
\end{align}
respectively, change the phase difference between the two partial waves
leaving the beam splitter.

\begin{table*}[tbh]
\caption{Quantum theoretical results for the probabilities of the quantum
walk after $l=1,\ldots,5$ steps (see Appendix~\ref{QT} for details on the calculation) for a particle initially localized at $x=0$.
The probabilities only depend on $\varphi_2$, not on $\varphi_1$.
For $l=1$ and $l=2$, the probabilities are identical to the ones of the SRW
(the first or only term in each column) which are given by $2^{-l} \binom{l}{(x+l)/2}$ if $x+l$
is even and are zero otherwise.
For more than two steps, the probabilities in each row exhibit a $\varphi_2 \to \varphi_2+\pi$ symmetry w.r.t.\ $x=0$.
Interference leads to the differences (red) between the probabilities of the SRW and the QW. The case $\varphi_1=\pi/2$ and $\varphi_2=-\pi/2$ is shown in Fig.~\ref{fig1}.}
\begin{ruledtabular}
\begin{tabular}{c|ccccccccccc}
     Step & \multicolumn{11}{c}{Lattice site (detector number) $x$}\\
     \hline
     $l$ &$ -5 $&$ -4 $&$ -3 $&$ -2 $&$ -1 $&$ 0 $&$ 1 $&$ 2 $&$ 3 $&$ 4 $&$ 5$\\
     \hline
     1 & $\phantom{-}0$ & $\phantom{-}0$ & $\phantom{-}0$ & $\phantom{-}0$ & $\phantom{-}\frac{1}{2}$ & $0$ & $\frac{1}{2}$ & $0$ & $0$ & $0$ & $0$ \\
     2 & $\phantom{-}0$ & $\phantom{-}0$ & $\phantom{-}0$ & $\phantom{-}\frac{1}{4}$ & $\phantom{-}0$ & $\frac{2}{4}$ & 0 & $\frac{1}{4}$ & 0 & $0$ & $0$ \\
     3 & $\phantom{-}0$ & $\phantom{-}0$ & $\phantom{-}\frac{1}{8}$ & $\phantom{-}0$ & $\phantom{-}\frac{3}{8}{\color{red}+\frac{2\cos\varphi_2}{8}}$ &
     $0$ & $\frac{3}{8}{\color{red}- \frac{2\cos\varphi_2}{8}}$ & $0$ & $\frac{1}{8}$ & 0 & $0$ \\
     4 & $\phantom{-}0$ & $\phantom{-}\frac{1}{16}$ & $\phantom{-}0$ & $\frac{4}{16}{\color{red}+\frac{2 + 4\cos\varphi_2}{16}}$ &
     $\phantom{-}0$ & $\frac{6}{16}{\color{red}-\frac{4}{16}}$ & $0$ &
     $\frac{4}{16}{\color{red}{\color{red}+\frac{2 - 4\cos\varphi_2}{16}}}$ & $0$ & $\frac{1}{16}$ & $0$ \\
     5 & $\phantom{-}\frac{1}{32}$ & $\phantom{-}0$ & $\frac{5}{32}{\color{red}+\frac{6 + 6\cos\varphi_2}{32}}$ & $\phantom{-}0$ &
     $\frac{10}{32}{\color{red}-\frac{6}{32}}$ &
     $\phantom{-}0$ & $\frac{10}{32}{\color{red}-\frac{6}{32}}$ & $0$ & $\frac{5}{32}{\color{red}+\frac{6 - 6\cos\varphi_2}{32}}$ & $0$ & $\frac{1}{32}$
   \end{tabular}
\end{ruledtabular}
\label{table_qw}
\label{tab1}
\end{table*}

Table~\ref{tab1} summarizes the theoretical results for the QW and the corresponding SRW.
For both types of walks, detectors with an odd (even) number $x$ will only register particles
if $l$ is also odd (even).
From the expressions in Table~\ref{table_qw} it also follows that the
probabilities to observe a particle do not depend on $\varphi_1$.
For more than two steps ($l>2$) the dependence on
$\varphi_2$ is sinusoidal, a characteristic feature of interference. Furthermore, the variance is larger than for the SRW and the peak of the distributions is not at the center anymore.

\begin{figure}[th]
   \centering
   \includegraphics[width=.8\hsize]{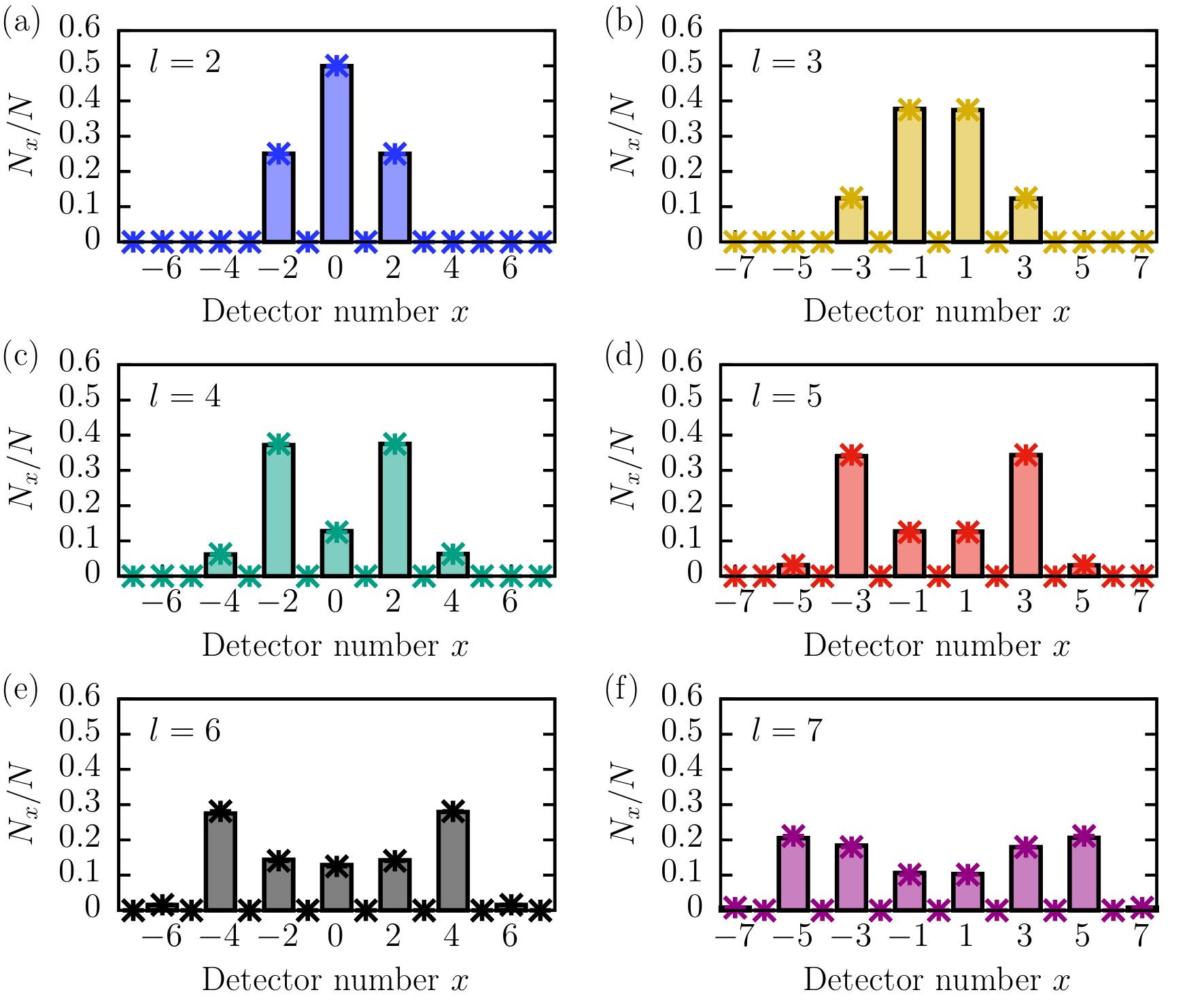}
   \caption{Results for the normalized number of detector counts $N_x/N$ as a
function of the detector number $x$, obtained by a DES of
the QW for $N=100\,000$ repetitions, $\varphi_1=\pi/2$ and
$\varphi_2=-\pi/2$ and for different numbers of steps $l=2,\ldots,7$, corresponding to (a)--(f).
The distributions from the DES (bars) match with the analytical results for the QW (asterisks, see Table~\ref{table_qw}).
For more than 3 steps, the distributions of the QW are broader than those of the SRW (see Table~\ref{table_qw}) because
of interference effects.
}
   \label{fig2}
\end{figure}

Implementing a DES for a network such as the one shown in Fig.~\ref{fig1} is straightforward.
We simply reuse, over-and-over again and without modification, the event-based algorithms that have been developed
to simulate the beam splitter, phase shifter, and detector~\cite{MICH14a} and connect
outputs to inputs of these algorithms strictly according to the diagram in Fig.~\ref{fig1}.
As the algorithms for all the different components
and the method to stitch them together have been discussed extensively and at great length elsewhere~\cite{MICH14a},
we omit the discussion of these aspects.
The reader interested in setting up her/his own DES should consult Ref.~\onlinecite{MICH14a} and papers cited therein.
Details of the implementation, specific for the application to QWs, can be found in Ref.~\onlinecite{NOCO16}. An example implementation in \textsc{python} is given in Appendix~\ref{Code} and available online~\cite{QuantumWalkOnline}.

Our implementation of the DES of the QW experiment shown in Fig.~\ref{fig1}
allows for more than $L=5$ levels ($X=11$ sites).
In general, the larger the number of beam splitters in the diagram,
the larger the number of particles has to be in order for the processors mimicking the beam splitters
to adapt sufficiently well to the ratio of particles arriving at the two input ports, i.e., representing the two sides at which photons can enter a beam splitter~\cite{RAED05b,MICH14a}.
Numerical experiments show that sending $N=100\,000$ particles through the network
is more than sufficient to go up to $L=7$ levels ($X=15$ sites) and to obtain data with good statistics.
Figure~\ref{fig2}(a)--(f) shows DES results after $l=2$ up to $l=7$ steps and for the phase shifts
$\varphi_1=\pi/2$ and $\varphi_2= -\pi/2$, as well as the results
obtained from the quantum-theoretical description (asterisks).
Other asymmetric cases are considered below and in Ref.~\onlinecite{NOCO16} and can be generated using the program given in Appendix~\ref{Code}.

The DES outcomes are in full agreement with the quantum-theoretical results.
In conclusion, the DES provides a local realist model that reproduces the quantum-theoretical results of the QW.

\section{Discrete-event simulation of a QW experiment with atoms~\cite{ROBE15}}\label{RE}

Robens \textit{et al.}\ experimentally implemented a four-level QW with cesium atoms
in a state-dependent optical potential~\cite{ROBE15}.
They made use of the fact that the two hyperfine states of the
electronic ground state of the cesium atom,
$\ket{F=4,\,m_F=4}$ (pseudo-spin up) and $\ket{F=3,\,m_F=3}$ (pseudo-spin down),
experience a different lattice potential~\cite{ROBE15}.
A microwave pulse can change the superposition of these two hyperfine states,
and the difference in sensitivity of the $\ket{F=4,\,m_F=4}$ and
$\ket{F=3,\,m_F=3}$ states to left- and right-handed polarized light can be used
to manipulate the position of the cesium atoms in the state-dependent potential~\cite{ROBE15}.

In the DES, a cesium atom with its two hyperfine states is represented by a particle carrying a two-state spin system.
Although we should not think of particles in the DES as objects observed in Nature,
to build a mental picture of what the DES is actually doing,
it may, for the present purpose, be very helpful to think of a particle and its spin
as a single photon and its polarization~\cite{ZHAO02}.
Therefore, and also for the uniformity of presentation,
we will formulate the DES model of the cesium-atom experiment
using the language of optics, using terms like beam splitters, phase shifters, etc.
As a matter of fact, as long as the dimension of the Hilbert space is finite,
it is always possible to reformulate the original problem
as a problem of photons traversing a network of optical components~\cite{RECK94} or,
equivalently, as a quantum gate circuit~\cite{NIEL00}.

\begin{figure}[th]
   \centering
   \includegraphics[width=0.9\textwidth]{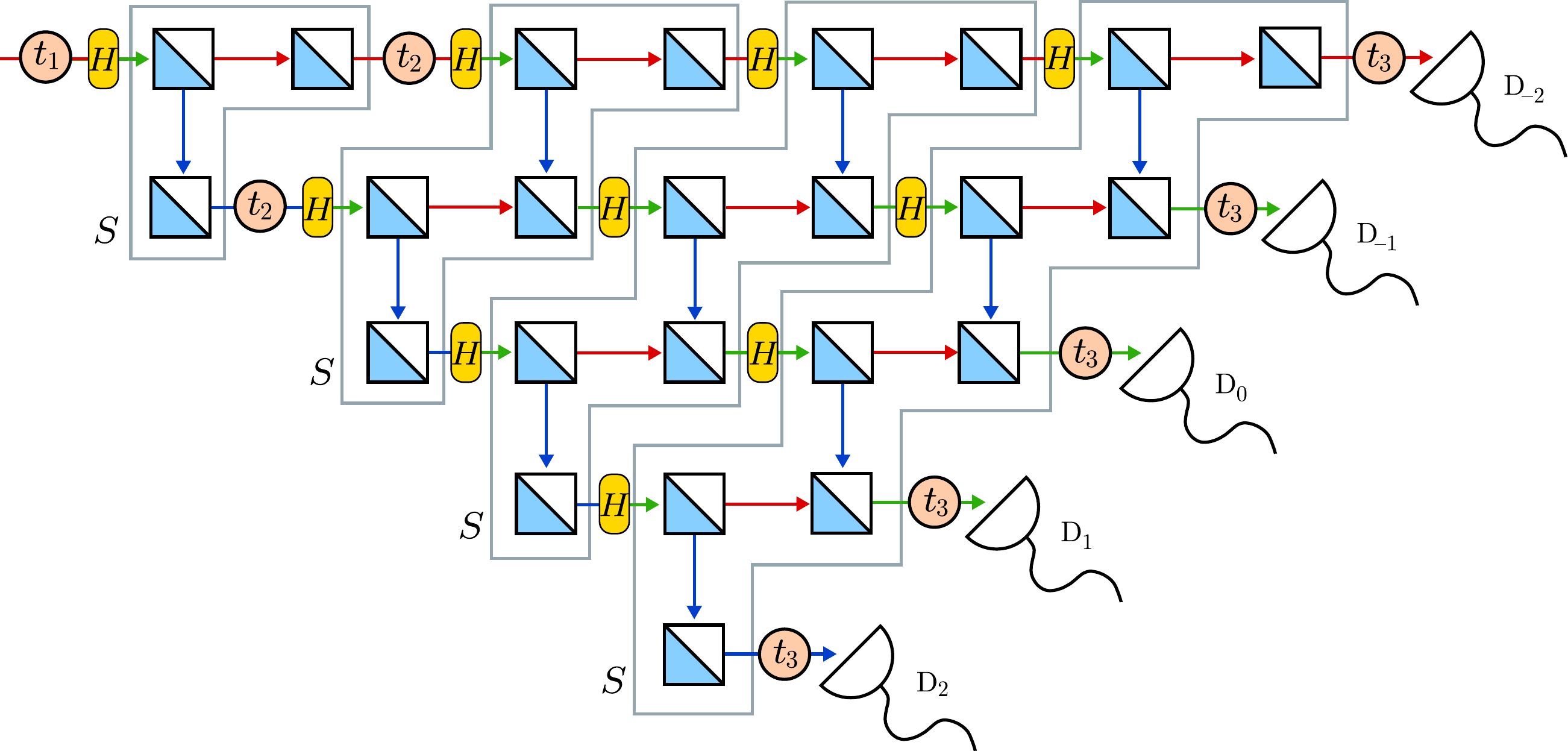}
   \caption{DES setup for the QW with polarized single photons of the same energy.
   Red horizontal (blue vertical) lines show the path of $h$ ($v$) polarized photons. Green lines represent the path of photons with a linear combination of $h$ and $v$ polarization.
   The input to the network consists of $h$ polarized photons only.
Square (blue/white) boxes represent polarizing beam splitters.
The oval (yellow) boxes perform a Hadamard transformation on the photon polarization.
Each gray region corresponds to a single-atom jump operation in the QW experiment~\cite{ROBE15}.
In quantum theory, the $H$ boxes correspond to Eq.~(\ref{INT3}) and the $S$ regions correspond to Eq.~(\ref{INT2}).
Circles with labels $t_1$, $t_2$, and $t_3$ denote the positions of the $Q(t_1)$, $Q(t_2)$, and $Q(t_3)$ measurements, respectively.
}
   \label{fig3}
\end{figure}

The basic ingredients of the DES are then the following~\cite{NOCO16}.
Distinguishing the cesium atoms on the basis of their hyperfine state
is implemented as the action of a polarizing beam splitter, separating $h$ and $v$ polarized photons (relative
to the entrance surface of the first polarizing beam splitter in Fig.~\ref{fig3}). In the context of the experiment, $h$ ($v$) corresponds to the hyperfine states $\ket{F=4,\,m_F=4}$ ($\ket{F=3,\,m_F=3}$).
The creation of the superposition of the hyperfine states
is realized by Hadamard transformations, i.e., a combination of half-wave plates and $\pi/2$
phase shifters~\cite{NOCO16,MICH14a}.
As $h$ and $v$ polarized photons do not interfere,
instead of the 50:50 beam splitters used in the DES of the QW model studied in section~\ref{QW},
we use polarizing beam splitters in order to let $h$ and $v$ polarized photons interfere~\cite{NOCO16}. A sample implementation of the DES is given in Appendix~\ref{Code}.

The ``photonics'' DES network that corresponds to the experiment with the cesium atoms~\cite{ROBE15}
is depicted in Fig.~\ref{fig3}.
Looking at Fig.~\ref{fig3}, it is easy to see that some of the polarizing beam splitters
(those that show only one input and one output line)
can be removed without affecting the operation of the network.
However, in our DES, we do not ``optimize'' the network for computational efficiency. As a matter of fact, the DES of the network in Fig.~\ref{fig3} is so fast that optimization is not worth the effort (the original implementation was written in \textsc{C++} \cite{NOCO16}, but even the demonstration in  \textsc{python} given in Appendix~\ref{Code} only takes a few seconds).

\begin{figure}[ht]
   \centering
   \includegraphics[width=.8\hsize]{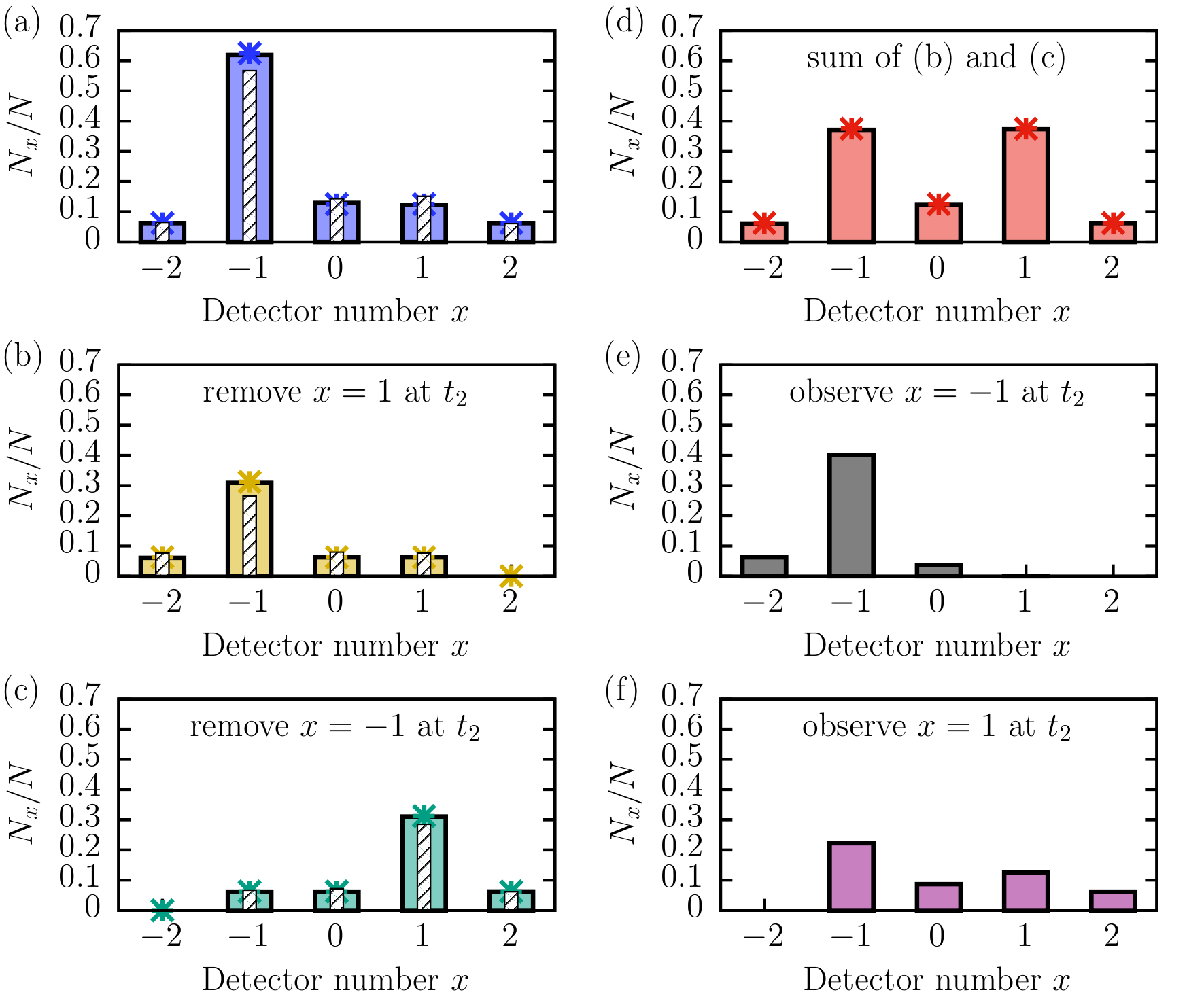}
   \caption{DES results (solid bars)
   of the normalized detector counts $N_x/N$ as a
function of the detector position $x$.
In each run, $N=100000$ particles were sent through the network shown in Fig.~\ref{fig3}. In (a)--(c), the solid bars represent the distribution where at position $t_2$, (a) no particles, (b) particles at $x=1$, (c) particles at $x=-1$ have been removed. In (d), the sum of (b) and (c) is shown to be symmetric and equal to the four-step QW shown in Fig.~\ref{fig2}(c) (for $x\mapsto x/2$). Panels (e) and (f) show the distributions resulting from only observing (and not removing) the particle at $t_2$. As their sum yields the distribution in (a), the observation does not affect the result and is thus non-invasive. Asterisks represent the ideal result obtained from quantum theory, i.e., $\sum_s|\langle 2x,s|(SH)^3|\psi\rangle|^2$ where $|\psi\rangle$ is given by (a) $SH|0,\uparrow\rangle$, (b) $|-1,\uparrow\rangle/\sqrt2$, or (c) $|+1,\downarrow\rangle/\sqrt2$ (see also Eqs.~(\ref{INT1})--(\ref{INT3})). There are no asterisks in (e) and (f) because this information is only accessible in the subquantum model.
The corresponding experimental data presented in Fig.~3 (a)--(c) of Ref.~\onlinecite{ROBE15} is (up to a normalization factor) indicated by the striped bars in panels (a)--(c).
}
\label{fig4}
\end{figure}

Figure~\ref{fig4}(a)--(c) shows that the DES reproduces the experimental results of Ref.~\onlinecite{ROBE15}. For convenience, the experimental data have been read off from Fig.~3(a)--(c) of Ref.~\onlinecite{ROBE15}, normalized, and plotted as striped bars in Fig.~\ref{fig4}(a)--(c).
Furthermore, we see that the DES produces the quantum-theoretical results
of the asymmetric four-step QW (asterisks).

The agreement between the DES and experimental data proves that, in contrast to the claim
made in Ref.~\onlinecite{ROBE15}, it is possible,  to describe a QW without a particle wave function,
but with particles following individual trajectories that are as well-defined as in the case of a SRW,
and local ``wave functions'' attached to each polarizing beam splitter~\cite{MICH14a}.
 We remark that the \texttt{learningrate} parameter of the beam splitters (see Appendix~\ref{Code}) can be used to tune the ``quantumness'' of the DES such that $\texttt{learningrate}=0$ yields the SRW and $0.9\le\texttt{learningrate}\le0.98$ yields the QW.

Obviously, the agreement between the DES and experimental data
seems to be in conflict with the common lore
that local realist models such as a DES cannot reproduce certain results of quantum physics.
It is therefore of interest to explore whether this conflict is fundamental or not. Recall that by construction, our DES model of the QW complies with the category 0
locality criteria, as defined in Ref.~\onlinecite{HESS19}.

Robens \textit{et al.}\ support their claim that the QW experiment
``rigorously excludes
(i.e., falsifies) any explanation of quantum transport based on classical, well-defined trajectories''
by demonstrating a violation of a LGI~\cite{ROBE15}
\begin{align}
   K = \langle Q(t_2)Q(t_1)\rangle + \langle Q(t_3)Q(t_2) \rangle - \langle Q(t_3)Q(t_1) \rangle \le 1
   ,
\label{ROB0}
\end{align}
where the $Q(t_i)$ are real numbers with $|Q(t_i)|\le 1$
and $t_i$ denote the position
at which the measurements are performed (equivalent to the time in the original formulation of the LGI).
We demonstrate, by means of a DES of their experiment, that their claim is unfounded.

\subsection{Procedure applied in the experiment~\cite{ROBE15}}
Robens \textit{et al.}\  set $t_1=0$ (initial state preparation $|\Phi^{(0)}\rangle=|x=0,{\uparrow}\rangle$,
start of a single-particle walk), $t_2=1$ (after the first single-atom jump),
and $t_3=4$ (after the fourth single-atom jump).
In Fig.~\ref{fig3}, each single-atom jump corresponds to a transition from one gray region to the next.
Circles with the labels $t_1$, $t_2$, and $t_3$ indicate the corresponding positions in the DES.
Robens \textit{et al.}\ proceed by choosing $Q(t_1)=Q(t_2)=1$ and assign $Q(t_3)=+1$
if at the fourth step, the particle is observed at $x>0$, and $Q(t_3)=-1$ otherwise~\cite{ROBE15}.
With these simplifications, Eq.~(\ref{ROB0}) reduces to
\begin{align}
   K = 1 + \langle Q(t_3)Q(t_2)\rangle - \langle Q(t_3)\rangle \le 1.
   \label{ROBE1}
\end{align}

In order to estimate $\langle Q(t_3) \rangle$,
Robens \textit{et al.}\ repeat the QW experiment about 400 times,
and compute the average of the measured $Q(t_3)$~\cite{ROBE15}.
To estimate $\langle Q(t_3)Q(t_2)\rangle$,
Robens \textit{et al.}\ need to repeat the same QW procedure two times in addition.
In the first (second) repetition, they measure the position at $t_2$, by what they believe is an ideal negative
measurement, and remove atoms that are measured at position $x=1$ ($x=-1$).
We cannot question the extent to which they really implemented
an ideal negative measurement in their experiment. In our DES of this experiment, however, it is trivial to perform an ideal negative measurement.
In both cases, the atoms continue their walk and are finally measured at $t_3$,
yielding either $Q(t_3)=-1$ or $Q(t_3)=+1$.
The average of the $Q(t_3)$'s is then denoted by
$\langle Q(t_3) \rangle_{x_2}$ where $x_2\in \{ -1,+1 \}$
indicates which atoms are kept at $t_2$.

With this data in hand, Robens \textit{et al.}\ compute the left-hand side of Eq.~(\ref{ROBE1}) as
\begin{align}
K = 1 + \sum\limits_{x_2=\pm 1}
P(x_2;\,t_2)\langle Q(t_3) \rangle_{x_2} - \langle Q(t_3)\rangle
,
\label{ROBE2}
\end{align}
where $P(x_2;\, t_2)$ denotes the probability that the atom was at position
$x_2=\pm1$ at $t_2$, the theoretical values being 1/2 (see the $l=1$ row of Table~\ref{tab1}).
Plugging in the experimentally obtained data, Robens \textit{et al.}\ find that~\cite{ROBE15}
\begin{align}
K = 1.435 \pm 0.074 > 1
,
\label{ROBE3}
\end{align}
and conclude that the ``reported violation of the LG inequality proves that the concept of a well-defined, classical trajectory is incompatible with the results obtained in a quantum-walk experiment.''~\cite{ROBE15}
This conclusion is unjustified, as we now show.

\subsection{Refutation of the claim}
Our demonstration consists of two steps.
First, we show that a DES of the QW performed with the same measurement procedure
as the one used by Robens \textit{et al.}\ reproduces their experimental results
and therefore also produces a violation of the LGI.
In this case, the DES also reproduces the results of the quantum-theoretical model in which we block the corresponding path labeled by $t_2$.
Second, because in a DES, performing non-invasive measurements is not an issue,
there is no need to perform three different runs to measure all the quantities
which appear in Eq.~(\ref{ROBE1}).
In fact, one DES run suffices to compute all the quantities that enter the LGI.
In this case, the DES also reproduces the quantum-theoretical results of the QW.

In the first step, we adopt the same procedure as in the real experiment~\cite{ROBE15},
namely we perform three DESs for a four-step QW.
In each DES run, the number of particles is $N=100000$.
In the first run, we compute $\langle Q(t_3)\rangle$  without removing particles at position $t_2$.
For the other two runs, at position $t_2$, we simulate an ideal negative measurement by removing the particles traveling
to the right ($h$ polarization) and downwards ($v$ polarization), respectively,
as Robens \textit{et al.}\ do in their experiment with the cesium atoms.

Direct confirmation that the DES reproduces the experimentally observed results
follows from comparing the data obtained using the removal process (see Fig.~\ref{fig4}(a)--(c))
with the corresponding data presented in Fig.~3(a)--(c) of Ref.~\onlinecite{ROBE15}.
Up to normalization factors, all results agree. Furthermore, the DES reproduces the quantum-theoretical results for the QW starting at $(t_1,x=0)$, $(t_2,x=-1)$, and $(t_2,x=+1)$, shown as asterisks in Fig.~\ref{fig4}(a)--(c), respectively.

Next, we compute $K$ as given in Eq.~(\ref{ROBE2}) from the data of the three different runs.
We estimate the statistical error on the value of $K$ by repeating the
three different runs ten times and obtain
\begin{align}
  K = 1.497 \pm 0.006 > 1,
  \label{ROBE4}
\end{align}
violating the LGI by several standard deviations.
In fact, the value of $K=1.497\pm 0.006$ is compatible
with the theoretical maximum violation of $K=1.5$,
achievable by this type of experiment~\cite{ROBE15}.

For the second step, we use the DES to perform truly ideal non-invasive ``measurements'' at $t_2$.
Instead of performing three DES runs (two of them removing certain particles),
we perform a single DES run, and only \emph{observe} the particle's position at $t_2$ (see Listing~\ref{codeexperiment} in Appendix~\ref{Code}).
We emphasize that in DES, this observation is truly non-invasive.

The resulting counts of the DES are shown in Fig.~\ref{fig4}(e) and (f).
From a comparison of Fig.~\ref{fig4}(b) and (c) with Fig.~\ref{fig4}(e) and (f), it is immediately clear
that there is a significant difference between the counts obtained by the three-run and single-run procedures. Furthermore, the distributions in Fig.~\ref{fig4}(e) and (f) add up to the original result in Fig.~\ref{fig4}(a). In contrast, the sum of the distributions in Fig.~\ref{fig4}(b) and (c), obtained by the invasive procedure, add up to the symmetric distribution in Fig.~\ref{fig4}(d), which is identical to the four-step QW shown in Fig.~\ref{fig2}(c).

The relevant question is whether Eq.~(\ref{ROBE1}) can still be violated. We compute $K$ from the data collected in a single DES of the QW and obtain
\begin{align}
   K = 0.999 \pm 0.002,
\end{align}
implying that there is no violation of Eq.~(\ref{ROBE1}) (up to statistical fluctuations).

The clear difference between results of the three-run and single-run procedure
proves that the violation of the LGI by the three-run procedure is not a property of the QW itself.
Instead, in the case at issue, the violation of the LGI is the result of using three different experimental scenarios with three different experimental data sets to compute the single quantity $K$.

It is worth mentioning that the data analysis used in other experiments that report violations of Bell-type inequalities shares similar features, in that correlations are computed from different subsets of a larger data set \cite{RAED16c}, which has been discussed in terms of the contextuality loophole~\cite{NIEU11}.
Such a procedure can, as Simpson's paradox nicely illustrates \cite{GRIM01}, lead to all kinds of interesting, paradoxical conclusions.

\section{Discussion and conclusion}
In this paper, we have proposed a subquantum model for quantum walks. The model is as realistic as the model for a simple
random walk and satisfies Einstein's criterion of locality, and uses
a digital computer and a discrete-event simulation
algorithm as a metaphor for realizable quantum walk experiments~\cite{JEON04,ROBE15}.
The subquantum model generates,
event-by-event, data that agrees with the quantum-theoretical
description of a quantum walk~\cite{JEON04}.

The subquantum model also reproduces the results of a quantum walk experiment with cesium atoms~\cite{ROBE15}.
In our simulation, the trajectories of each individual particle can be followed. Therefore, the conclusion made in Ref.~\onlinecite{ROBE15} ``that the concept of a well-defined, classical trajectory is incompatible with the results obtained in a quantum-walk experiment'' is unjustified. The results presented in this paper can be reproduced with the \textsc{python} programs provided in Appendix~\ref{Code} and online~\cite{QuantumWalkOnline}.

Our subquantum model based on discrete-event simulation can reproduce the experimental data of quantum walk experiments as well as many other optics and neutron-interferometry experiments~\cite{RAED05b,MICH14a,RAED16c,NOCO16}.
This suggests that standardized software that allows for simulations of single events observed in (quantum) physics experiments may lead to a new kind of theory.
Whether the discrete-event simulation approach can be modified/generalized to attain the descriptive power of a theory, formulated in terms of software (i.e., a well-defined set of rules stated in terms of a programming language) rather than in the conventional language of theoretical physics, is a challenging project for future research.

Being a realistic and Einstein-local model, a salient feature of our simulation approach
is the absence of concepts such as particle-wave duality, Born's rule, and other
concepts which are characteristic of quantum theory.
Regarding the foundations of the latter, it is of interest to mention that one of the rules by which the
discrete-event simulation operates requires
attaching a kind of ``local wave function'' to some of the event-based processing units (such as the beam splitters)~\cite{RAED05b,MICH14a}.
This is very reminiscent of a proposal by Duane, who showed that one can explain
the diffraction of X-rays from a crystal without reference to interference of waves,
by adding, to the quantum rules for energy and angular momentum, a similar rule for the linear momentum~\cite{DUAN23}.
In essence, Duane suggested that instead of invoking the particle-wave character, for model building
it may be more effective to let particles (not waves) interact with the
multitude of wave-like motion that is already present in the crystal~\cite{LAND65}.
As we have shown in this paper, this idea can be combined with discrete-event simulation
to yield a local realist model for a quantum walk.

\appendix
\section{Quantum theoretical description of the QW}\label{QT}

In this appendix, we outline the quantum theoretical description of the QW that can be used to obtain the theoretical results given in Table~\ref{table_qw}. Using the same notation
as in Eqs.~(\ref{INT0}) and~(\ref{INT2}), we denote the basis states as $|x,s\rangle$,
where $x\in\{-L,\ldots,L\}$ and $s\in\{\uparrow,\downarrow\}$. In the context of
Fig.~\ref{fig1}, $x$ increases from the top-right corner to the bottom-left corner
as indicated by the indices of the detectors. The label $s=\uparrow$ ($s=\downarrow$) corresponds to the horizontal (vertical) arrows in Fig.~\ref{fig1}.

With this notation, the optical elements at position $x$ in Fig.~\ref{fig1} are described in terms of the operators
\begin{align}
    B^{(x)} &= |x\rangle\!\langle x|\otimes M_\mathrm{BS} = \frac 1 {\sqrt 2} (|x,\uparrow\rangle+i|x,\downarrow\rangle)\langle x,\uparrow\!| + \frac 1 {\sqrt 2} (i|x,\uparrow\rangle+|x,\downarrow\rangle)\langle x,\downarrow\!|,\\
    P_1^{(x)} &= |x\rangle\!\langle x|\otimes M_{\varphi_1} = e^{i\varphi_1} |x,\uparrow\rangle\!\langle x,\uparrow\!| + |x,\downarrow\rangle\!\langle x,\downarrow\!|,\\
    P_2^{(x)} &= |x\rangle\!\langle x|\otimes M_{\varphi_2} = |x,\uparrow\rangle\!\langle x,\uparrow\!| + e^{i\varphi_2}|x,\downarrow\rangle\!\langle x,\downarrow\!|,
\end{align}
where the beam-splitter matrix $M_\mathrm{BS}$ is given in Eq.~(\ref{MBS}) and the phase-shifter matrices $M_{\varphi_1}$
and $M_{\varphi_2}$ are given in Eq.~(\ref{MPhi1Phi2}). Combining the three operators, we obtain the description of a dashed box at position $x$ in Fig.~\ref{fig1} as
\begin{align}
    T^{(x)} = P_2^{(x)}B^{(x)}P_1^{(x)} = \frac 1 {\sqrt 2} |x\rangle\!\langle x| \otimes
    \begin{pmatrix} e^{i\varphi_1} & i \\ ie^{i(\varphi_1+\varphi_2)} & e^{i\varphi_2}
    \end{pmatrix}.
\end{align}
The particle jump between the dashed boxes in Fig.~\ref{fig1} is described by the shift operator $S$ given by Eq.~(\ref{INT2}) which shifts the $\uparrow$ state from $x$ to $x-1$ and the $\downarrow$ state from $x$ to $x+1$.
For the initial state $|\Phi^{(0)}\rangle=|0,{\uparrow}\rangle$, we thus obtain the state after $l=1,2,\ldots$ steps as
\begin{align}
    |\Phi^{(l)}\rangle =
    \begin{cases}
     S B^0 |0,{\uparrow}\rangle & (l=1) \\
     S (T^{(-l+1)} + T^{(-l+3)} + \cdots + T^{(l-1)}) |\Phi^{(l-1)}\rangle & (l>1) \\
    \end{cases}.
\end{align}
The probabilities in Table~\ref{table_qw} at position $x$ after $l$ steps (corresponding to the detector $D_x$ in Fig.~\ref{fig1}) can then be computed as
\begin{align}
    p(x,l)=|\langle x,\uparrow\!|\Phi^{(l)}\rangle|^2+|\langle x,\downarrow\!|\Phi^{(l)}\rangle|^2.
\end{align}

\section{\textsc{python} implementation of the subquantum model}\label{Code}

In Listings~\ref{codegeneral}--\ref{codedes}, we give a simple \textsc{python} implementation of the DES used for the QWs studied in this paper (the three source files are also available online~\cite{QuantumWalkOnline}). The code shows that DES provides a local realist model in which the trajectory of each individual particle can be followed. Nonetheless, the collective statistics of the individual particles perfectly agree with the experimental results and the quantum-theoretical predictions (see Figs.~\ref{fig2} and~\ref{fig4}).

\lstdefinestyle{codestyle}{
    language=python,
    commentstyle=\color[rgb]{0.13,0.54,0.13},
    keywordstyle=\color{blue},
    numberstyle=\tiny\color{red},
    stringstyle=\color[rgb]{0.7,0.0,0.0},
    basicstyle=\ttfamily\scriptsize,
    emph={beamsplitter,__init__},
    breakatwhitespace=false,
    breaklines=true,
    captionpos=t,
    keepspaces=true,
    numbersep=5pt,
    showspaces=false,
    showstringspaces=false,
    showtabs=false,
    tabsize=2,
    frame=single
}
\lstset{style=codestyle}

\begin{lstlisting}[label=codegeneral,caption=\texttt{qw\_general.py}: General QW implementing the network in Fig.~\ref{fig1} and used to obtain the results in Fig.~\ref{fig2}]
#!/usr/bin/env python3
import numpy as np
from des_elements import particle,beamsplitter

L = 4            # number of steps/levels in the quantum walk
N = 100000       # number of individual particles simulated
phi1 = np.pi/2   # phase shifter angle 1
phi2 = -np.pi/2  # phase shifter angle 2

# create local beamsplitters at each level l and position x
beamsplitters = {}
for l in range(1,L+1):
    beamsplitters[l] = {}
    for x in range(-l+1,l,2):
        beamsplitters[l][x] = beamsplitter(polarizing=False)

# create local detectors (particle counters) at level L
detectors = {}
for x in range(-L,L+1,2):
    detectors[x] = 0

# simulate each individual particle's trajectory through the network in Fig. 1
for n in range(N):
    p = particle()  # create new particle
    x = 0           # current position of the particle on its trajectory
    port = 0        # current port at the beam splitters (0: left/right, 1: top/down)
    x,port = beamsplitters[1][x].passthru(p,x,port)  # first beam splitter at level 1
    for l in range(2,L+1):                           # each dashed box in Fig. 1
        if port == 0: p.apply_phaseshift(phi1)           # phase shifter 1
        x,port = beamsplitters[l][x].passthru(p,x,port)  # beam splitter
        if port == 1: p.apply_phaseshift(phi2)           # phase shifter 2
    detectors[x] += 1  # finally detect the particle at position x

# print normalized detector counts for Fig. 2
print('# x Nx/N (for l='+str(L)+')')
for x,Nx in detectors.items():
    print(x, Nx/N)
\end{lstlisting}
\begin{lstlisting}[label=codeexperiment,caption=\texttt{qw\_experiment.py}: QW implementing the network in Fig.~\ref{fig3} and used to obtain the results shown in Fig.~\ref{fig4}]
#!/usr/bin/env python3
import numpy as np
from des_elements import particle,beamsplitter

N = 100000       # number of individual particles simulated
remove = 'none'  # which particles to remove at t2 ('none', 'x=1', 'x=-1')

# local beamsplitters and detectors (x grows from top-right to bottom-left in Fig. 3)
beamsplitters = {
    1: {x: beamsplitter() for x in [-1, 0, 1]},                       # 1. gray region
    2: {x: beamsplitter() for x in [-2, -1, 0, 1, 2]},                # 2. gray region
    3: {x: beamsplitter() for x in [-3, -2, -1, 0, 1, 2, 3]},         # 3. gray region
    4: {x: beamsplitter() for x in [-4, -3, -2, -1, 0, 1, 2, 3, 4]},  # 4. gray region
}
detectors = {
    'all':  {x: 0 for x in [-2, -1, 0, 1, 2]},  # count all particles
    'x=1':  {x: 0 for x in [-2, -1, 0, 1, 2]},  # count those that went along x=1 at t2
    'x=-1': {x: 0 for x in [-2, -1, 0, 1, 2]},  # count those that went along x=-1 at t2
}

# simulate each individual particle's trajectory through the network in Fig. 3
for n in range(N):
    p = particle()  # create new particle
    x = 0           # current position of the particle on its trajectory
    port = 0        # current port at the beam splitters (0: left/right, 1: top/down)
    for r in [1,2,3,4]:  # pass through each gray region r in Fig. 3
        p.apply_hadamard()                               # the H box in Fig. 3
        x,port = beamsplitters[r][x].passthru(p,x,0)     # always enter at port 0 (left)
        _,port = beamsplitters[r][x].passthru(p,x,port)  # second beam splitter in r
        if port != 0:  # during learning, a few particles may leave at the wrong port
            break
        if r == 1:
            p.x_at_t2 = x  # at t2, tag particle for "ideal non-invasive measurement"
            if remove == 'x=1'  and x == 1:  break
            if remove == 'x=-1' and x == -1: break
        if r == 4:
            detectors['all'][x/2] += 1  # at t3, the particle triggers a detector
            if p.x_at_t2 ==  1: detectors['x=1'][x/2]  += 1
            if p.x_at_t2 == -1: detectors['x=-1'][x/2] += 1

# print normalized detector counts for Fig. 4 (second column depends on 'remove')
print('# x Fig.4(a)-(c) Fig.4(e) Fig.4(f)')
for x in [-2,-1,0,1,2]:
    print(x, detectors['all'][x]/N, detectors['x=-1'][x]/N, detectors['x=1'][x]/N)
\end{lstlisting}
\begin{lstlisting}[label=codedes,caption=\texttt{des\_elements.py}: Definition of the basic DES elements]
#!/usr/bin/env python3
import numpy as np
from numpy import sin,cos,exp,sqrt,pi
from numpy.random import rand

class particle:
    def __init__(self, xi=0): # xi: polarization (h: xi=0, v: xi=pi/2)
        self.m = np.array([   # every particle carries a message m
            sin(xi),  # vertical component (v)
            cos(xi)   # horizontal component (h)
        ])
    def apply_phaseshift(self, phi):
        self.m = exp(1j*phi) * self.m
    def apply_hadamard(self):
        self.m = 1/sqrt(2.) * np.array([
            [1, 1],
            [1, -1]
        ]) @ self.m

class beamsplitter: 
    def __init__(self, polarizing=True, learningrate=0.98):
        if polarizing:  # polarizing beam splitter (transmits h and reflects v)
            self.TBS = np.array([
                [0, 1j, 0,  0],
                [1j, 0, 0,  0],
                [0,  0, 1, 0],
                [0,  0, 0, 1]
            ])
        else:  # normal 50:50 beam splitter (splits both h and v)
            self.TBS = 1/sqrt(2.) * np.array([
                [1,  1j,  0,  0],
                [1j,  1,  0,  0],
                [0,   0,  1, 1j],
                [0,   0, 1j,  1]
            ])
        self.lr = learningrate
        # the state of each local beam splitter is determined by 6 numbers
        self.X = [.5, .5]
        self.Y = [[1, 0], [0, 1]]
    def passthru(self, p, x, i):  # p: particle, x: position, i: input port (0/1)
        self.Y[i] = p.m
        self.X[i]   = self.lr * self.X[i] + (1 - self.lr)
        self.X[1-i] = self.lr * self.X[1-i]
        W = self.TBS @ [
            sqrt(self.X[0])*self.Y[0][0],
            sqrt(self.X[1])*self.Y[1][0],
            sqrt(self.X[0])*self.Y[0][1],
            sqrt(self.X[1])*self.Y[1][1],
        ]
        # determine output port and update particle message
        prob0 = abs(W[0])**2 + abs(W[2])**2
        if rand() < prob0:
            p.m = np.array([W[0], W[2]]) / sqrt(prob0)
            return x-1, 0  # return new position and output port
        else:
            p.m = np.array([W[1], W[3]]) / sqrt(1-prob0)
            return x+1, 1  # return new position and output port
\end{lstlisting}

\bibliography{bibliography_merged}
\end{document}